# Resonant Auger spectroscopy at the $L_{2,3}$ shake-up thresholds as a probe of electron correlation effects in nickel

M. Magnuson, N. Wassdahl, A. Nilsson, A. Föhlisch, J. Nordgren and N. Mårtensson

*Department of Physics, Uppsala University, P. O. Box 530, S-751 21 Uppsala, Sweden*

**Abstract**

The excitation energy dependence of the three-hole satellites in the $L_3$-$M_{4,5}M_{4,5}$ and $L_2$-$M_{4,5}M_{4,5}$ Auger spectra of nickel metal has been measured using synchrotron radiation. The satellite behavior in the non-radiative emission spectra at the $L_3$ and $L_2$ thresholds is compared and the influence of the Coster-Kronig channel explored. The three-hole satellite intensity at the $L_3$ Auger emission line reveals a peak structure at 5 eV above the $L_3$ threshold attributed to resonant processes at the $2p^53d^9$ shake-up threshold. This is discussed in connection with the 6-eV feature in the x-ray absorption spectrum.

## 1 Introduction

The 3$d$ transition metal systems have been extensively studied because these systems show influence of many-body effects which are interesting both from an experimental and theoretical point of view [1,2,3,4,5,6,7]. The occurence of distinguishable satellite structures in the spectra is a sign of localization tendencies of the 3$d$ valence electrons which retain some of their atomic-like properties in the metal. In this respect, Ni is often considered a prototype system in the 3$d$ transition metal series regarding strong correlations and configuration interaction (CI) in the ground, core- and valence-excited states. Ni has recieved much attention especially in connection with the well-known 6-eV photoemission satellite of both the valence-band and core-levels [8,9,10,11]. The strong satellite structures and the small bandwidth in the Ni metal are directly connected to and give important information on the correlation effects which have been described in terms of a complex, energy dependent self-energy [12,13,14,15,16]. The 6-eV satellite in valence band spectra has been observed to display Fano-like intensity variations both at the 3$p$ [17] and 2$p$ [18] core-level thresholds due to interference effects. A good understanding of the interesting physical properties of Ni requires detailed spectroscopic studies of the electronic structure performed at sufficiently high resolution.

The interpretation of the 6-eV feature in the Ni $L_{2,3}$ x-ray absorption spectrum is still controversial. It has been observed that the photon energies of the main line and the 6 eV feature in the x-ray absorption spectrum coincide with the corresponding core level binding energies in the x-ray photoelectron spectrum [19]. The satellite in the photoemission spectrum has been found to be due to a $2p^53d^9$ double hole state and it has therefore been natural to identify the x-ray absorption feature with the same type of final state.

In this contribution, we address the controversial issue of the interpretation of the 6-eV feature at the $L_{2,3}$ edges in the x-ray absorption spectrum in Ni based on a quantitative study of the three-hole satellites ($3d^7$ final state configuration) at the $L_{2,3}$ shake-up thresholds. The large $L_{2,3}$ spin-orbit splitting makes it possible to compare the behavior at the two thresholds separately, thereby exploring the consequences of the competing Coster-Kronig (CK) channel at the $L_2$ threshold. The relative weight of the $3d^7$ final state satellite intensity of the non-radiative decay spectra provide a quantitative probe of the population of the localized $2p^53d^9$ core-level shake-up states in the Ni $L_{2,3}$ absorption spectrum. This gives direct insight into the origin of the 6-eV spectral feature in the absorption spectrum which has been discussed both in terms of delocalized one-electron band states [20] and localized $2p^53d^9$ multiplet states [21].

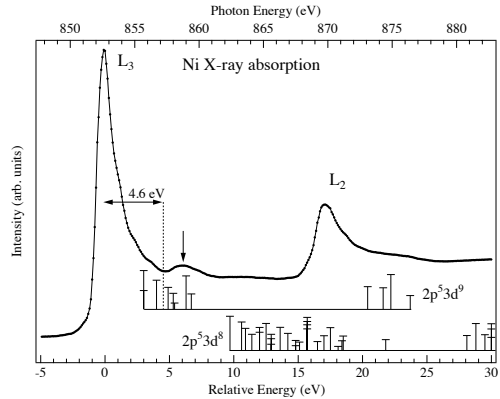

**Figure 1:** The $L_{2,3}$ x-ray absorption spectrum of Ni. The energy positions of the localized $2p^53d^9$ and $2p^53d^8$ shake-up configurations are shown at the bottom.

## 2 Experimental Details

The Auger measurements were performed at beamline 8.0 at the *Advanced Light Source* (ALS), Lawrence Berkeley National Laboratory (LBNL). The beamline comprises a 5 cm period undulator and a spherical-grating monochromator. The experimental station built at Uppsala University includes a rotatable Scienta SES200 electron spectrometer [22]. The base pressure was lower than $2\times10^{-10}$ Torr during preparations and measurements. The Ni(100) single crystal sample was of high purity and crystal quality, and cleaned by means of cyclic argon-ion bombardment and annealing to remove surface contaminants. The sample was oriented so that the photons were incident at about $7^o$ grazing angle with the polarization vector of the x-rays in the plane of the sample. The electron spectrometer was oriented near the sample normal and perpendicular to the photon beam. This geometry increases the weight of the Auger matrix element relative to the direct photoemission matrix element. During the Auger measurements the resolution of the monochromator of the beamline was 0.15 eV. The Ni $L_{2,3}$ absorption spectrum was measured in normal incidence at beamline 10-1 at the *Stanford Synchrotron Radiation Center* (SSRL) using a SGM monochromator with a resolution of 0.10 eV. The sample was made of a 200 Å thick Ni film grown on Si(100) with a 100 Å Ru buffer layer and a 20 Å Ru capping layer to prevent oxidation.

## 3 Results and Discussion

Figure 1 shows an x-ray absorption spectrum of the Ni $L_{2,3}$ region. The main peaks at 852.7 eV and 870.0 eV on the photon energy scale at the top are due to resonant excitation to the $2p^5(P_{3/2})3d^{10}$ ($L_3$) and $2p^5(P_{1/2})3d^{10}$ ($L_2$) core-excited states, respectively. The 6-eV feature is indicated by the arrow.

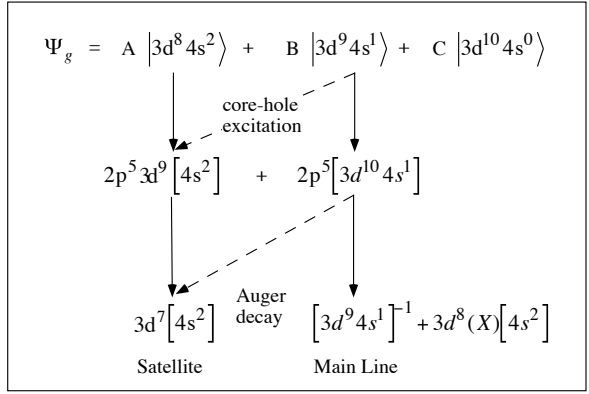

**Figure 2:** A schematic diagram of the atomic multiplet configurations in Ni, and the excitation and Auger decay channels for the main line and the $3d^7$ final state. A ≈ 15-20 %, B ≈ 60-70 %, and C ≈ 15-20 % [10, 21].

For the discussion of the different decay processes in Ni metal, for probing the different core-hole states, we use the localized atomic-like approach. In order to explicitly account for the multiplet effects, atomic Dirac-Fock calculations [4,23] for the localized $2p^53d^9$ and $2p^53d^8$ core-level shake-up states are shown at the bottom of Fig. 1 relative to the $L_3$ XPS energy [19] which represents the Fermi level for the unoccupied valence states in the presence of the core hole. The calculated intensities of the multiplet levels were assumed to attain intensities according to their multiplicities ($2J+1$). The center-of-gravity of the $2p^53d^9$ multiplets are located at 4.6 eV relative energy both for the $L_3$ and $L_2$ shake-up configurations [4].

Figure 2 is a schematic excitation and decay diagram of the main $2p$ core-hole excitations and Auger decay channels in Ni. It has been proposed that the ground state ($\Psi_g$) of Ni can be expressed by the superposition of $3d^8$, $3d^9$ and $3d^{10}$ configurations with relative weights of 15-20%, 60-70% and 15-20%, respectively [21].

For simplicity, we denote the ground state in Ni as $[3d^94s^1]$, where the square brackets indicate that the valence electrons are in the delocalized metallic state. Normal Auger processes in Ni are known to lead to localized (atomic-like) as well as delocalized (band-like) final states. These are the same types of final states that appear in the direct valence band photoemission [24,25]. The two final states in normal valence band photoemission processes can be written as $[3d^94s^1]^{-1}$ and $3d^8(X)[4s^2]$, for the delocalized (well screened) $3d$ band and the localized (poorly screened) 6-eV satellite, respectively, where $X$ is the spectroscopic assignment of the atomic $3d^8$-configuration. In resonant photoemission there is an interplay between direct photoemission and Auger-like (autoionization) processes [10, 18]. When the measurements are performed at sufficiently high resolution, it is in many cases possible to distinguish between Auger-like and photoemission-like processes in the non-radiative decay spectra [18]. Here, the core-level binding energy defines the onset of the regular Auger processes. These processes involve the decay of well-defined states of a given energy, leading to Auger emission energies which are independent of the excitation energy.

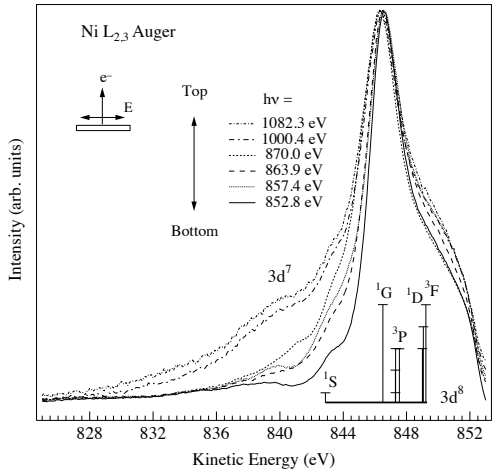

**Figure 3:** A series of $L_{2,3}$ Auger spectra measured with excitation energies ranging from 852.8 eV up to 1082.3 eV. The localized $3d^8$ multiplet configuration at the $L_3$ threshold is indicated at the bottom of the figure.

For higher photon energies, excitations from the $[3d^8 4s^2]$ ground state to the $2p^5(P_{3/2})3d^9[4s^2]$ core-level shake-up configuration are energetically allowed and provide the localized $3d^7[4s^2]$ final-state satellite channel in the Auger spectra. At excitation energies above the $L_2$ threshold, CK decay from the $2p^5(P_{1/2})[3d^{10}4s^1]$ to the $2p^5(P_{3/2})3d^9[4s^2]$ core-excited configuration is also possible. The dashed lines in Fig. 2 indicate possible initial and final state shake-up events between the $3d$ and $4s$ subshells.

Figure 3 compares a few Ni $L_{2,3}$ Auger spectra measured at various photon energies, from 852.8 eV at the $L_3$ threshold, up to energies as high as 1082.3 eV, far above both the $L_2$ ($E_B$=870.0 eV) and the $L_1$ ($E_B$=1008.6 eV) thresholds. The kinetic energy is given relative to the Fermi level and the excitation energy thus corresponds to the high-energy cutoff in each spectrum. The spectra are shown after subtraction of an integral Shirley background [26] and at the bottom the localized $3d^8$ multiplet configuration is indicated. For excitation energies above the $L_3$ threshold ($E_B$=852.7 eV), a rapidly growing $3d^7$ satellite tail extending towards lower kinetic energies from the main line is observed in the spectra.

Starting with the resonant spectrum at the $L_3$ threshold at the bottom in Fig. 3, mainly the $2p^5(P_{3/2})[3d^{10}4s^1]$ core-hole has been created in the excitation. We observe two types of final states, band-like states within 2.3 eV from the Fermi level and split-off atomic-like $3d^8$ final states 6 eV from the Fermi level. The atomic-like $3d^8$ multiplets consists of the $^1G$ term at the main peak and the $^3F$, $^3P$ and $^1D$ terms which overlap with the band-like states, while the $^1S$ term can be distinguished on the low-energy side of the spectrum. A small satellite contribution can be observed even in the threshold spectrum between 835 and 840 eV kinetic energy [11] due to final state shake-up events between the $3d$ and $4s$ subshells (dashed lines in Fig. 2). In the 857.4 eV and 863.9 eV spectra, the $2p^5(P_{3/2})3d^9[4s^2]$ configuration is excited which opens up the satellite decay channel producing localized $3d^7$ final states on the low kinetic energy side [27]. In Fig. 3 it can also be noted that the satellite intensity is indeed higher at 857.4 eV than at 863.9 eV excitation energy.

In the spectrum excited at 870.0 eV, a $2p^5(P_{1/2})[3d^{10}4s^1]$ core-hole is created which decays into the $[3d^94s^1]^{-1}$ and $3d^8(X)[4s^2]$ final states in the energy region of the $L_2$ Auger peak at about 865 eV kinetic energy. However, the $2p^5(P_{1/2})[3d^{10}4s^1]$ core-hole can also decay via a CK transition into the $2p^5(P_{3/2})3d^9[4s^2]$ configuration, which subsequently can decay into a $3d^7$ satellite final state. It is interesting to note the appearance of a new feature in the $3d^7$ satellite spectrum at 841 eV kinetic energy as the photon energy reaches or exceeds the $L_2$ threshold. We propose that this difference in the shape of the satellite spectrum is due to a different relative population of the $2p^53d^9$ multiplet states for the shake-up and the CK processes. The distinct onset of the 841 eV feature is evident from a sequence of spectra recorded with photon energy steps of 0.5 eV (not shown here). The present experimental accuracy and energy tuning range enables us to make a clear distinction between the shake-up and the CK induced satellite structures which was earlier more limited in these respects[5]. At the $L_2$ threshold, $2p^5(P_{3/2})[3d^84s^2]$ core-hole states are also created, producing $3d^6$ satellite final states at even lower kinetic energies. More of the $3d^6$ satellite final states are produced in the spectrum excited at 1000.4 eV, between the $L_2$ and $L_1$ thresholds. In the spectrum excited at 1082.3 eV well above the $L_1$ threshold, the $3d^6$ satellite final states are produced either directly from the $2s$ core holes, or through the $2s3d^9$ core holes.

The Auger process is normally described as a "two-step" process. In the first step a photoelectron is excited from a core-orbital. In the decay step the core hole is filled by a valence electron and an Auger electron is emitted from the valence levels. With separated excitation and emission steps, the satellite contribution can be separated from the main line by a subtraction procedure. Figure 4 shows the relative integrated satellite intensity at the $L_3$ and $L_2$ Auger emission lines (in percent), normalized to the satellite-free threshold spectra excited at 852.8 eV and 870.0 eV, respectively. The intensities were extracted after Shirley background subtraction and normalization by subtracting the satellite-free threshold spectra so that the intensities of the difference spectra were always positive. For excitation energies above the $L_3$ threshold, a rapid increase of the relative satellite intensity is observed 3-5 eV above the $L_3$ threshold to a maximum value of about 15 %. For excitation energies approaching the $L_2$ threshold, a step of rapid intensity increase is observed up to a new plateau at about 25 %. This intensity step shows that the CK process plays an important role for the development of the satellite structure. Above the $L_2$ threshold, the intensity is again slowly increasing due to additional shake-up and shake-off processes up towards the $L_1$ level, where additional $L_1$ CK processes are enabled.

The peak structure observed at about 5 eV above the $L_3$ threshold in Fig. 4 corresponds well to the $2p^5(P_{3/2})3d^9$ core-level shake-up threshold. The various multiplet terms in this configuration have different onsets ranging from 3 eV to about 7 eV, as shown in Fig. 1. The 5-eV peak represents a sum over the different intensity contributions originating from the different terms in the $2p^5(P_{3/2})3d^9$ configuration. The satellite intensity variation indicates that the multiplets around 3-5 eV contribute more to the intensity than those located at 5-7 eV. This implies that the strongest multiplet terms of the of the $2p^5(P_{3/2})3d^9$ configuration have onsets lower than 6 eV which suggests that the 6-eV feature in the x-ray absorption spectrum (indicated by the arrow in Fig. 1) has another origin than from localized multiplet states. It has also been observed in calculations of ionic Ni compounds having a dominating $3d^8$ configuration in the ground state that the strongest multiplets are located at the lowest energies while those at higher energies have only a small contribution [28]. However, for metallic Ni the matrix elements could be somewhat different.

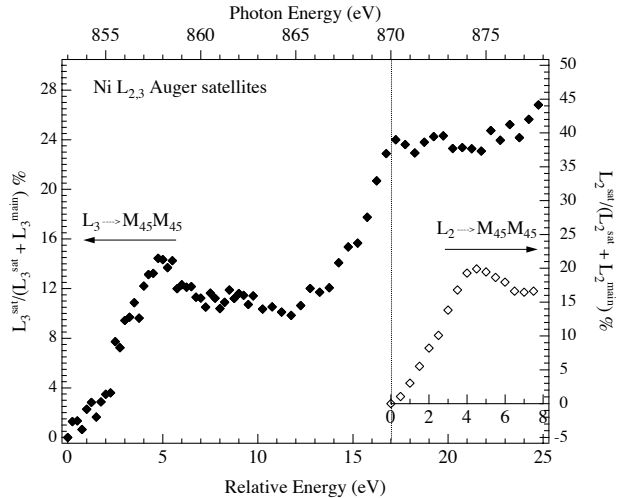

**Figure 4:** The integrated $3d^7$ satellite intensity vs. excess photon energy above the $L_3$ and $L_2$ thresholds.

The satellite intensity at the $L_2$ threshold also shows a similar behavior as at the $L_3$ threshold with a peak maximum at about 4.5 eV above threshold. This peak is related to the $2p^5(P_{1/2})3d^9$ configuration in a similar way as at the $L_3$ threshold, although the relative intensity is different. It should also be noted that the satellite intensity at the $L_3$ peak increases significantly already at 12 eV (5 eV below the $L_2$ edge) which is an indication that $2p^5(P_{3/2})3d^8$ core-level double shake-up states are excited.

It has been found that the 6-eV feature in the absorption spectrum can be reproduced by one-electron band-structure calculations [20,29,30] which is an indication that the origin of this feature could be due to the hybridized $s-p$ band. Chen *et al.* [31,32] observed that the 6-eV feature vanished in magnetic x-ray circular dichroism (MCD) spectra. Instead a new feature appeared at 4 eV in the dichroic difference spectrum which could not be reproduced in the one-electron model [20]. This feature was instead attributed to correlation effects in the absorption final state, implying the presence of both $3d^8$ and $3d^9$ atomic configurations in the ground state [21].

The presence of resonant structures at 5 eV above the $L_{2,3}$ edges implies that both local $2p^53d^9$ multiplet configurations and band-structure effects are essential features in the x-ray absorption spectrum. The qualitative difference between the 6 eV spectral feature in the $2p\rightarrow 3d$ x-ray absorption spectrum and the 6 eV satellite in the valence- and core-photoemission spectra of Ni is to a large extent due to the difference in the matrix elements involved in these different types of processes. In other words, the 6 eV feature in the absorption spectrum is due to a critical point in the unoccupied band structure at this energy[20, 33] and the atomic multiplet configurations are less important for the shape of the x-ray absorption spectrum.

# 4 Conclusions

The excitation energy dependence of Auger spectra of nickel metal has been measured and the $3d^7$ three-hole satellites close to the $2p_{3/2,1/2}$ core-level thresholds have been investigated. For excitation energies above the $2p_{1/2}$ edge, a satellite peak in the Auger spectra appears as a consequence of the population of the $2p^5(P_{3/2})3d^9$ core-excited states due to the additional Coster-Kronig channel. For excitation energies close to the $2p$ shake-up thresholds, the $3d^7$ final state satellite is also found to have a different intensity distribution. The integrated satellite intensity reveals a maximum at 5 eV above threshold. This intensity variation provides a direct probe of the spin-integrated population of the core-excited $2p^53d^9$ states, i. e., the intermediate states which retain a localized character. The 6-eV feature in the x-ray absorption spectrum can be explained within the one-electron band-structure picture while the strength of $2p^53d^9$ local multiplet states has a maximum at somewhat lower energy as revealed by the resonant Auger spectra.

# 5 Acknowledgments

This work was supported by the Swedish Natural Science Research Council (NFR), the Göran Gustavsson Foundation for Research in Natural Sciences and Medicine and the Swedish Institute (SI). ALS is supported by the U.S. Department of Energy, under contract No. DE-AC03-76SF00098.